\documentclass[english,twocolumn,prb,superscriptaddress,showpacs]{revtex4}
\usepackage[T1]{fontenc}
\usepackage[latin1]{inputenc}
\usepackage{graphicx}
\usepackage{color}
\usepackage{babel}

\begin{document}

\title{Analysis of the electric field gradient in the perovskites
  SrTiO$_3$ and BaTiO$_3$: density functional and model calculations}

\author{K.\ Koch}

\affiliation{Max-Planck-Institute for Chemical Physics of Solids,
  01187 Dresden, Germany}

\author{R.O.\ Kuzian}

\affiliation{Institute for Problems of Materials Science
  Krzhizhanovskogo 3, 03180 Kiev, Ukraine}

\author{K. Koepernik}

\affiliation{Leibniz Institute for Solid State and Materials Research,
  01171 Dresden, Germany}

\author{I.V.\ Kondakova}

\affiliation{Institute for Problems of Materials Science
  Krzhizhanovskogo 3, 03180 Kiev, Ukraine}

\author{H.\ Rosner}

\affiliation{Max-Planck-Institute for Chemical Physics of Solids,
  01187 Dresden, Germany}

\begin{abstract}
We analyze recent measurements [R.\ Blinc, V.\ V.\ Laguta, B.\ Zalar,
  M.\ Itoh and H.\ Krakauer, J.\ Phys.\ : Condens.\ Matter
  \textbf{20}, 085204 (2008)] of the electric field gradient on the
oxygen site in the perovskites SrTiO$_{3}$ and BaTiO$_{3}$, which
revealed, in agreement with calculations, a large difference in the
EFG for these two compounds.  In order to analyze the origin of this
difference, we have performed density functional electronic structure
calculations within the local-orbital scheme FPLO. Our analysis yields
the counter-intuitive behavior that the EFG increases upon lattice
expansion. Applying the standard model for perovskites, the effective
two-level $p$-$d$ Hamiltonian, can not explain the observed
behavior. In order to describe the EFG dependence correctly, a model
beyond this usually sufficient $p$-$d$ Hamiltonian is needed.  We
demonstrate that the counter-intuitive increase of the EFG upon
lattice expansion can be explained by a $s$-$p$-$d$ model, containing
the contribution of the oxygen 2$s$ states to the crystal field on the
Ti site.  The proposed model extension is of general relevance for all
related transition metal oxides with similar crystal structure.
\end{abstract}

\date{\today}


\maketitle 

\section{Introduction}

Perovskite compounds $AB$O$_{3}$, with $A$ being an alkali, alkaline
earth or rare earth metal and $B$ a transition metal element, attract
much attention because of their importance both for fundamental
science and for technological applications \cite{LinesGlass}. Although
the high-temperature cubic phase has a very simple crystal structure,
this does not prevent these compounds to exhibit a large variety of
physical properties rendering the perovskites to model compounds for
studies of a large variety of different physical phenomena. Within the
perovskite family, we find superconductivity, e.g. in
K$_x$Ba$_{1-x}$BiO$_3$ \cite{Matth88}, giant magnetoresistance,
e.g. in LaMnO$_3$ \cite{Moritomo96}, orbital ordering, e.g. in
YTiO$_3$ \cite{Ishihara02} and ferroelectricity, e.g. in BaTiO$_{3}$
\cite{Cohen92}. The latter phenomena are of large interest because of
technological applications.

The compounds SrTiO$_{3}$ (STO) and BaTiO$_{3}$ (BTO) are usually
considered to be isovalent. The valence and conduction bands of the
two perovskites are formed by $p$-states of oxygen and $d$-states of
titanium. In the high-temperature cubic phase, the Ti and O
sub-lattices have the identical geometry for STO and BTO, the lattice
parameters being $a$=3.8996~\AA{} \cite{STO} and $a$=4.009~\AA{}
\cite{LinesGlass} respectively. As the temperature lowers, both
compounds experience a softening of an optical phonon mode, which
corresponds to Ti motion towards the oxygen \cite{LinesGlass}. BTO
exhibits a succession of phase transitions, from the high-temperature
cubic perovskite phase to ferroelectric structures with tetragonal,
orthorhombic and rhombohedral symmetry \cite{LinesGlass}. In contrast,
STO behaves as an incipient ferroelectric in the sense that it remains
paraelectric down to the lowest temperatures, exhibiting nevertheless
a very large static dielectric response. It undergoes an
antiferrodistortive phase transition at 105~K to a tetragonal
($I4/mcm$) phase, but this transition is of non-polar character and
has little influence on the dielectric properties\cite{Sai2000}.

The first determination of the $^{17}$O electric field gradient (EFG)
on the oxygen site in perovskites was recently reported for STO and
BTO \cite{Blinc08} together with first-principle calculations using
the linearized augmented plane wave (LAPW) method was used. The most
striking feature in the experimental and theoretical data is the large
difference of the EFGs between the two compounds.  The calculational
investigation of Blinc {\it et al.}  concluded, that the magnitude of
the EFG of $^{17}$O in BTO is larger than the EFG of $^{17}$O in STO
due to two effects: $(i)$ larger lattice parameters in BTO compared to
STO and $(ii)$ a larger ionic radius of Ba compared to Sr.  While the
experimental determination (NMR) can not provide the sign of the EFG,
the LAPW calculation yielded a negative EFG. A negative EFG
corresponds to a prolate electron density, which implies 
the importance of covalence effects.

In order to elucidate the origin of the sign of and the different
contributions to the EFG, we have performed first-principle
calculations using a local orbital code (FPLO \cite{FPLO}) that is
especially suited to address these questions due to its representation
of the potential and the density allowing easy decomposition.  The
calculational details of our investigation are given in
Sec.~\ref{calmeth}, and the obtained results are presented in
Sec.~\ref{Far}. These results can not be explained by intuitive
models, which are also described in this section. Therefore, a more
complex microscopic model Hamiltonian is introduced in
Sec.~\ref{disc}.  Using the properties of this $p$-$d$ like
Hamiltonian, an agreement with the obtained experimental and
theoretical results and a deeper, microscopically based understanding
is obtained.

\section{Calculation methods}\label{calmeth}

The electronic band structure calculations were performed with the
full-potential local-orbital minimum basis code FPLO (version 5.00-19)
\cite{FPLO} within the local density approximation.  In the scalar
relativistic calculations the exchange and correlation potential of
Perdew and Wang \cite{PW} was employed.  As basis sets Ba
(4d5s5p/6s6p5d+4f7s7p), Sr (4s4p/5s5p4d+6s6p), Ti (3s3p3d/4s4p4d+5s5p)
and O (2s2p3d+3s3p) were chosen for semicore/valence+polarization
states. The high lying states improve the completeness of the basis
which is especially important for accurate EFG calculations. The lower
lying states were treated fully relativistic as core states. A well
converged $k$-mesh of 455 $k$-points was used in the irreducible part
of the Brillouin zone.

\section{FPLO analysis results}\label{Far}

\begin{table}
\begin{tabular}{|c|r|r|c|}
\hline 
 & SrTiO$_3$ & BaTiO$_3$& Ref. 
\\
\hline
$|V_{zz}^{exp}|$& 1.62 \ & 2.46 \ & Ref.~\onlinecite{Blinc08} 
\\
\hline
$V_{zz}^{cal}$& -1.00 \  & -2.35 \ &Ref.~\onlinecite{Blinc08} 
\\
\hline
$V_{zz}^{cal}$& 1.00 \ & 2.44 \ &Eq.~(\ref{totEFG})
\\
\hline
$V_{zz}^{on}$ &-0.21 \ & 1.39 \ &Eq.~(\ref{Vonsite})
%$V_{zz}^{on,net}$ &-0.19 \ & 1.64 \ &Eq.~()
\\
\hline
$V_{zz}^{off}$& 1.21 \ & 1.05 \ &Eq.~(\ref{Voffsite})
\\
\hline
$V_{zz,pp}^{on,net}$& 96\%& 107\%&Eq.~(\ref{eq:Vzzpp})
\\
\hline
\end{tabular}

\caption{The experimental and calculated values of the EFG (in 10$^{21}$
V/m$^{2}$) on the oxygen site in the cubic phase of the two perovskites. The
last four lines refer to equations given in the appendix.}
\label{Tab1}
\end{table}

In FPLO, the EFG on a nucleus at a given lattice site may be
represented as the sum of two contributions: An on-site contribution
$V_{zz}^{on}$ (see Eq.(\ref{Vonsite})), which comes from the on-site
contribution of the electron density of the given lattice site, and a
second term, the off-site contribution $V_{zz}^{off}$ (see
Eq.(\ref{Voffsite})), which results from the potential of all other
atoms (see App.~A). The on-site contribution $V_{zz}^{on}$ can be
analyzed further.  It can be split up in $p$-$p$, $s$-$d$ and $d$-$d$
contributions (see App.~B).

\begin{figure} [htbp]
\includegraphics*[width=\columnwidth]{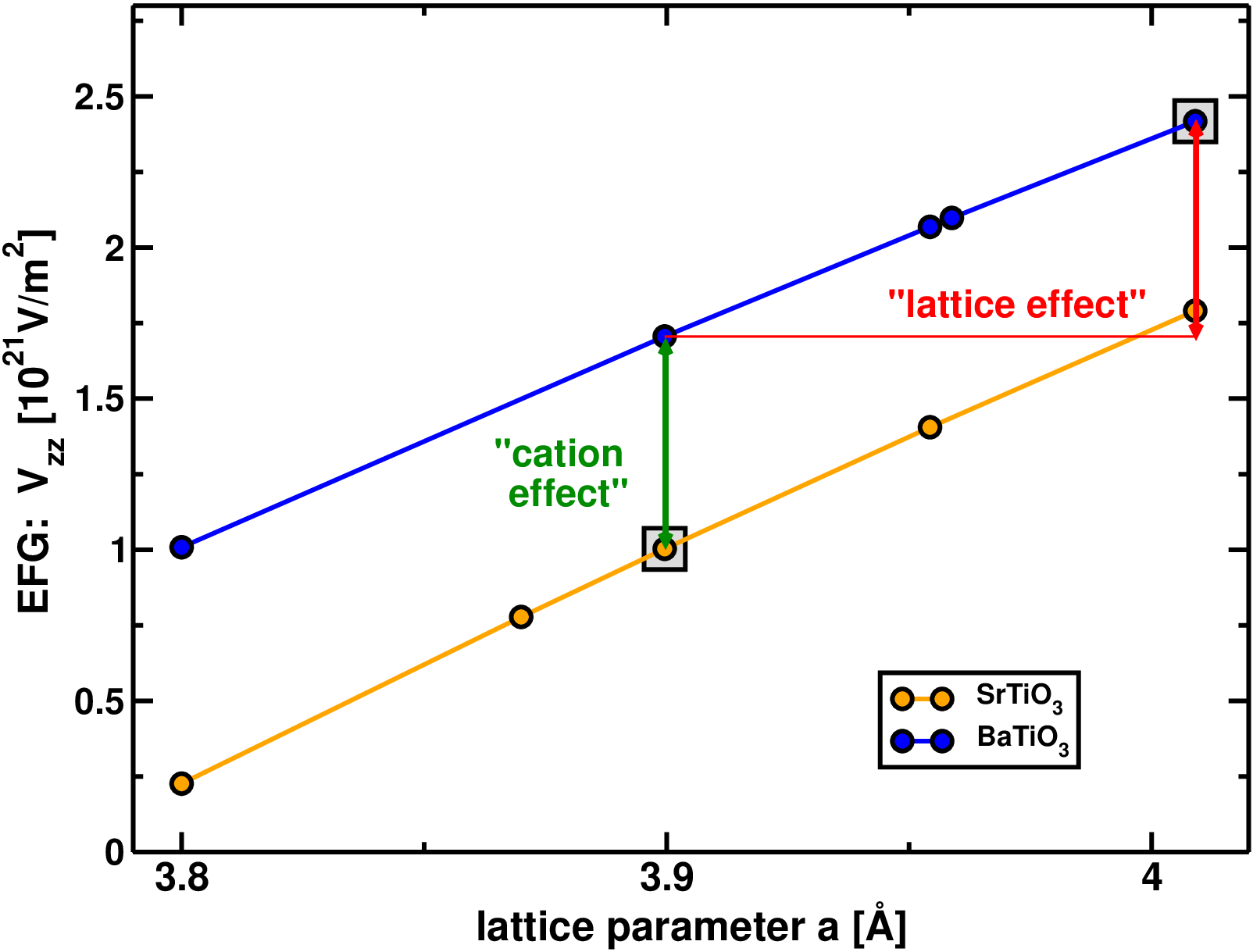}
\caption{Calculated $V_{zz}$ in dependence of the lattice parameter
  $a$. $V_{zz}$ for the experimental lattice parameter is marked by a
  shaded square. The ``cation'' and ``lattice effect'', which are
  responsible for the difference in $V_{zz}$ for these two compounds
  are indicated by the red and black arrow, respectively.  {\bf
    Inset:} The anisotropy count $\Delta p$ (see text) in dependence
  of the lattice parameter $a$.}
\label{Figure1}
\end{figure}

The on- and off-site contributions as well as their sum and the
dominating $p$-$p$ contribution (see Eq.~(\ref{eq:Vzzpp})) are shown
in Tab.~\ref{Tab1}. Whereas the total EFG for $^{17}$O in BTO agrees
well with the experiment (1~\% deviation), the total EFG for $^{17}$O
in STO is in discrepancy with the experiment (38~\% deviation), see
Tab.~\ref{Tab1}. Compared to the EFGs calculated with the LAPW code in
Ref.~\onlinecite{Blinc08}, we obtain almost the same absolute value of
$V_{zz}$ but the opposite sign, see Tab.~\ref{Tab1}. Our calculated
EFGs as a function of the lattice parameter $a$ for both compounds
reveal the same tendency as observed in Ref.~\onlinecite{Blinc08}: The
absolute value of the EFG {increases} under the lattice expansion (see
Fig.~\ref{Figure1}).  From Fig.~\ref{Figure1} we also conclude that
the EFG of BTO is not only larger than the EFG of STO due to larger
lattice parameters (``lattice effect''), but also due to an ``cation
effect'', which is responsible for the remaining difference. This
lattice effect is demonstrated by the shift between the two EFG curves
in Fig.~\ref{Figure1}.

The increase of the (absolute value of the) EFG upon lattice expansion
is rather counter-intuitive. In the traditional approach, the
spherically symmetric electronic shell of an ion is perturbed by the
potential of the external (point) charges of the solid. As a result,
the total EFG on the ion nucleus is caused by the EFG of the external
potential, and is roughly proportional to it. It is clear that this
approach predicts {the opposite} tendency: The strength of the
external potential is inversely proportional to the lattice constant
and thus the (absolute value of the) EFG should diminish under the
lattice expansion. The failure of this approach to describe the
observed behavior of the EFG indicates that a fully ionic description
of the perovskites is inappropriate.

In an alternative approach, the electronic shell of the atom is
disturbed by the hybridization of the wave functions with the states
of the surrounding atoms. The hybridization results in the asymmetry
of the electronic cloud of the atom and the EFG on its
nucleus. Apparently, this covalent approach predicts the same tendency
as the ionic one: It is usually believed that the hybridization
diminishes with the increase of the bond length. In both approaches we
may say: When expanding the lattice, we diminish its influence on the
atom, and the electronic shell should become closer to that of the
free atom. Hence, we come to the conclusion: The (absolute value of
the) EFG should diminish under the lattice expansion, which is
opposite to the experimental observation and the results of both
first-principle calculations.  We will tackle this problem in detail
in Sec.~\ref{disc}.

Another problem it the different sign of the EFG obtained from the
two different band structure codes. If the sign of the EFG is taken
into account, the slope in our graph (Fig.~\ref{Figure1}) is opposite
to the slope in the graph obtained with the LAPW code (Fig.~5 in
Ref.~\onlinecite{Blinc08}). Since the NMR experiment is not sensitive to the
sign of the EFG, we will investigate the influence of the lattice
expansion on the different contributions to the EFG to get more
insight in this issue.

\begin{figure} [tbp]
\includegraphics*[width=\columnwidth]{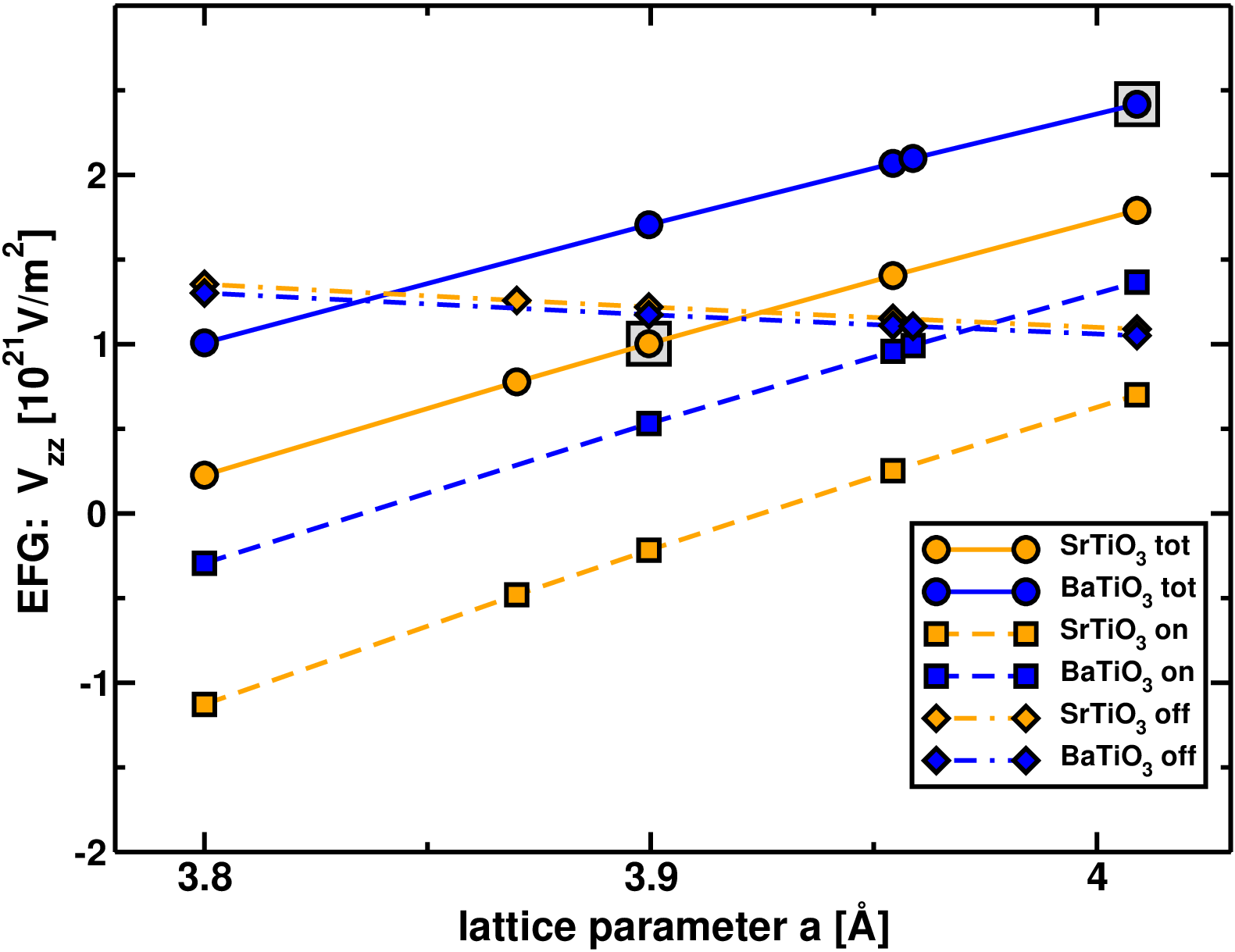}
\caption{The on-site $V_{zz}^{on}$, off-site $V_{zz}^{off}$ and total
  EFG as a function of the lattice parameter $a$. The grey shaded
  squares mark the experimental lattice parameter for $V_{zz}$.}
\label{Figure2}
\end{figure}

Our calculations show that both the on-site and the off-site
contribution to the EFG have comparable values for the perovskite
lattice, see Tab.~\ref{Tab1} and Fig.~\ref{Figure2}. In
Fig.~\ref{Figure2}, the two contributions, $V_{zz}^{on}$ (dashed line)
and $V_{zz}^{off}$ (dash-point line) and the total EFG (full line) are
shown.  Whereas the off-site EFG decreases only slightly upon lattice
expansion, the on-site EFG increases strongly with increasing lattice
parameters, resulting in the significant increase of the total EFG. We
also observe that the off-site EFG is almost identical for these two
structures, which is in line with the observed weak dependence of
$V_{zz}^{off}$ on the lattice parameters.

The on-site EFG is mainly caused by electrons with $p$ character, see
table \ref{Tab1}. Therefore, we will investigate the corresponding
anisotropy count $\Delta p$ \cite{Blaha88}.  In the perovskite
structure $AB$O$_{3}$, the oxygen site has axial symmetry, and the
$z$-axis is directed along the $B$-O bond. Thus, the anisotropy count
is the difference between the population of the oxygen 2$p$ $\sigma$-
(corresponds to $p_{z})$ and $\pi$- (corresponds to
$p_{x,y})$ orbitals.  In the inset of Fig.~\ref{Figure1} we see that the
anisotropy count $\Delta p$ increases with the lattice expansion. This
is in agreement with the increasing on-site EFG.  If we focus on BTO,
where the experimental and calculated (for the experimental lattice
parameter $a=4.009$~\AA) value for the EFG agree very well, we see
that this positive $V_{zz}$ corresponds to a positive $\Delta p$. That
means the $p$ electron density (responsible for the EFG) has an oblate
shape, since more electrons are occupying the $p_{x,y}$-orbitals than
the $p_{z}$-orbital, which is in agreement with the positive sign of
the EFG. 

\begin{figure} [tbp]
\includegraphics*[width=\columnwidth]{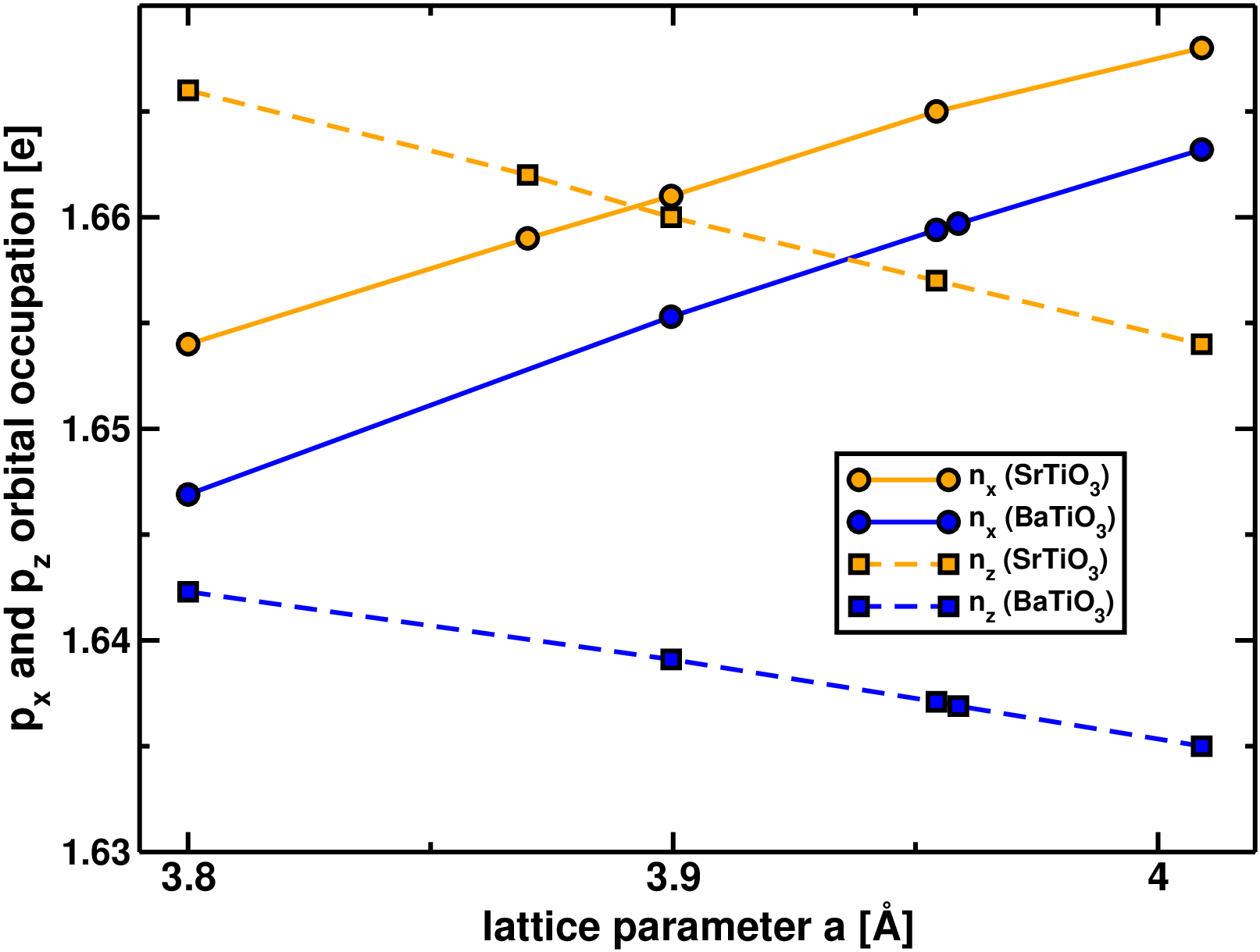}
  \caption{Occupation of $p_x$ and $p_z$ states in dependence of the
    lattice parameter $a$.}
\label{Figure3}
\end{figure}

After concluding that the sign of $V_{zz}$ for $^{17}$O for both STO
and BTO should be positive, we come back to the counter intuitive
behavior of the increasing EFG upon lattice
expansion. Fig.~\ref{Figure3} reveals that the increase of $\Delta p$
under lattice expansion, which is responsible for the increasing EFG
upon lattice expansion, is due to an increasing occupation of $pi$-
(corresponds to $p_{x,y})$ and an decreasing population of $\sigma$-
(corresponds to $p_{z})$ orbitals.

\section{Discussion}\label{disc}

In order to understand this anomalous behavior of the
$\sigma$-orbital, we will analyze the main features of the electronic
structure of perovskites.  Detailed band structure studies of
perovskite compounds were performed by Mattheiss
\cite{Matth69,Matth72,Matth70}, who also proposed a first
tight-binding fit for the band dispersions. Wolfram \textit{et al.}
\cite{Wolfram72,Wolfram73,Wolfram82} (cf. also
Ref.~\onlinecite{Pros87}) developed a very simple model (Wolfram and
Ellialtioglu, WE) for the valence and conduction bands, which reflects
their basic properties. The WE model includes the $d$-orbitals of the
$B$ ion and the $p$-orbitals of the oxygen. Wolfram \textit{et al.}
pointed a quasi-two-dimensional character of the bands out, which is
due to the symmetry of the orbitals. If one retains only nearest
neighbor hoppings, the total $14\times14$ Hamiltonian matrix (five
$d$-orbitals and 9 $p$-orbitals) acquires block-diagonal form
at every value of the momentum. The three $3\times3$ matrices describe
the $\pi_{ij}$-bands ($ij=xy,yz,xz$). Every $d_{ij}$-orbital of the
$t_{2g}$ symmetry couples with its own combination of oxygen $2p$
$\pi$-orbitals, which lie in the same plane perpendicular to the bond
direction. They form a pair of bonding and anti-bonding states.  The
remaining combination of the $2p$ $\pi$-orbitals in the same plane
forms the non-bonding band. Wolfram \textit{et al.} call this group of
bands $\pi$-bands. The states described by the $5\times5$ block matrix
are called $\sigma$-bands, since they are formed by oxygen $2p$
$\sigma$-orbitals, which are coupled with the $e_{g}$
($d_{x^{2}-y^{2}}$ and $d_{z^{2}}$) orbitals of the B ion. This matrix
decouples into one non-bonding band and two pairs of bonding and
anti-bonding bands.

\begin{figure}[t]
\begin{center}
 \includegraphics[clip,width=9cm]{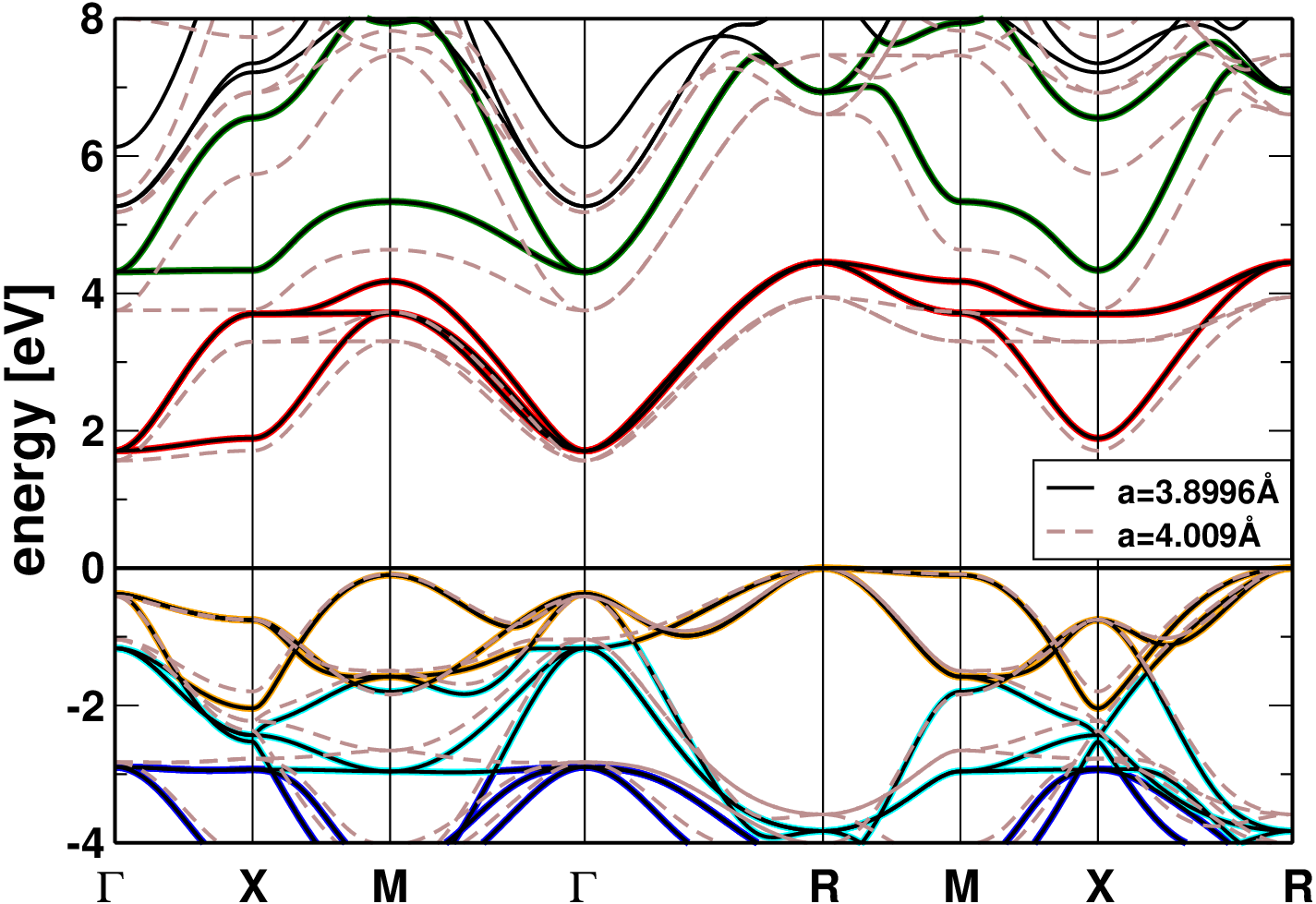}
\end{center}
\caption{SrTiO$_{3}$: band structure for two different lattice
  parameters $a=3.8996$~{\AA} (black/colored full lines) and
  $a=4.009$~{\AA} (brown dashed lines). The different band characters
  are given by different colors: blue (bonding, $\sigma$), cyan
  (bonding, $\pi$), orange (non-bonding), red (anti-bonding,
  $\pi_{ij}$) and green (anti-bonding, $d_{eg}$), see text. Since it
  is not easy to interpret the valence band, the colors in the valence
  band are only approximate.}
\label{Figure4} 
\end{figure}

Fig.~\ref{Figure4} shows the calculated band structure for STO for two
different lattice parameters $a$. The features mentioned above are
clearly seen (cf.~Fig.~2 of Ref.~\onlinecite{Wolfram82}). The
anti-bonding $\pi_{ij}$-bands are situated between 2 and 4 eV, where
the $\pi_{yz}$-band is almost dispersionless in the direction
$\Gamma\rightarrow X$.  This manifests the quasi-two-dimensional
character of the bands. The bands originating from the $d$
$e_{g}$-orbitals are in the range from 4 to 8 eV, where the band
expressing $d_{z^{2}}$ character is dispersionless along the
$\Gamma\rightarrow X$ direction. The valence band has a more complex
character due to additional mixing from the direct $p-p$ hopping. This
is neglected in the simple version of the WE model. Nevertheless, we
see that the non-bonding bands lie on top of the valence band and
have a much smaller dispersion than the bonding bands, which lie below
$-1$~eV ($\pi_{ij}$) and below $-3$~eV ($\sigma$-bands). The latter
have a larger dispersion due to much larger $d-p$ hoppings.

Although the Kohn Sham theory is not good for excitation spectra, or
obtaining the correct energy gap, it yields reliable occupation
numbers, on-site energies and transfer integrals, especially in the
absence of strong correlations. Therefore, we can use our LDA band
structure to obtain reliable parameters as input for further treatment
using model Hamiltonians.

In the following, we explore within the WE model how the occupation numbers
and the resulting anisotropy count for the $p$-orbitals depend on the
lattice parameters.  In dielectric compounds like STO and BTO, the
bonding and non-bonding states are fully occupied. Contrary to the
non-bonding bands, which have almost pure $p$-character, the bonding
and anti-bonding bands are mixed $p$-$d$-bands. The population of the
$p$-orbitals is given by the sum of the occupation numbers of the
non-bonding and the bonding bands, whereof the latter are lattice
parameter dependent.

Every pair of bonding and anti-bonding states is described by an effective
two-level model \cite{Wolfram73} 
\begin{eqnarray}
\nonumber
\hat{H}_{m}
&=&\Delta_{m}\left(\left|d,\mathbf{k}\right\rangle \left\langle d,\mathbf{k}\right|
-\left|p,\mathbf{k}\right\rangle \left\langle p,\mathbf{k}\right|\right)
\\&&
+V_{m}f_{m\mathbf{k}}\left(\left|d,\mathbf{k}\right\rangle \left\langle p,\mathbf{k}\right|
+\left|p,\mathbf{k}\right\rangle \left\langle d,\mathbf{k}\right|\right).
\label{eq:Htls}
\label{eq:Htlsproj}
\end{eqnarray}
Here, $m$ describes the character of the band $m=\pi,\sigma$ and
$f_{m\mathbf{k}}$ is a dimensionless function, which depends on the
dimensionless variable $\mathbf{k}a$ (note that $\mathbf{k}$ is
measured in units of $\pi/a$, so neither $\mathbf{k}a$ nor
$f_{m\mathbf{k}}$ depends on $a$). The state mixing is defined by the
interplay of the on-site energy difference $\Delta_{m}$ and the
transfer integral $V_{m}$, which determines the bandwidth of the
corresponding band.  The eigenstates of the Hamiltonian
Eq.~(\ref{eq:Htls}) have the form
\begin{eqnarray}
\left|\mathbf{k},\nu\right\rangle
=c_{d\mathbf{k}\nu}\left|d,\mathbf{k}\right\rangle
+c_{p\mathbf{k}\nu}\left|p,\mathbf{k}\right\rangle,
\label{eq:eigen}
\end{eqnarray} and for the $p$-states,  the following energies and
occupation numbers are obtained
\begin{eqnarray}
\label{Vvonhier}
E_{\mathbf{k}m\nu} & = &
  \nu\sqrt{\Delta_{m}^{2}+V_{m}^{2}\left(f_{m\mathbf{k}}\right)^{2}},\ 
\nu=\pm1\label{eq:Eknu}
\\ 
n_{p\mathbf{k}m\nu}
  & \equiv&
2\left|c_{p\mathbf{k}\nu}\right|^{2}= 
  1-\Delta_{m}/E_{\mathbf{k}\nu}.\label{eq:Nknu}
\\ 
n_{d\mathbf{k}m\nu} 
  &\equiv&
2\left|c_{d\mathbf{k}\nu}\right|^{2}= 
  2\left(1-\left|c_{p\mathbf{k}\nu}\right|^{2}\right)
=1+\frac{\Delta_{m}}{E_{\mathbf{k}\nu}}
\label{eq:Ndknu}
\end{eqnarray}
Here, $\nu=+1$ describes the anti-bonding and $\nu=-1$ describes the
bonding band. In this two-level system, two asymptotic behaviors are
possible.  First, $\Delta_{m}/V_{m}\rightarrow\infty$, which yields
for the occupation numbers of the bonding bands
$n_{p\mathbf{k}m,-1}\rightarrow2$ and
$n_{d\mathbf{k}m,-1}\rightarrow0$. In this case, both electrons are in
the $p$-state of the ligand ion and $d$-states are empty, called the
ionic limit.
Second, $\Delta_{m}/V_{m}\rightarrow0$, which yields for the occupation
numbers $n_{p\mathbf{k}m,-1}\rightarrow1$ and
$n_{d\mathbf{k}m,-1}\rightarrow1$.  In this case, the electrons are equally
shared by the $p$-, and $d$-states. This is the covalent limit.
From the trends in Fig.~\ref{Figure3}, we observe that while the
population of the $p_\pi$-orbitals increases, the population of the
$p_\sigma$-orbitals decreases. This means the Ti-O $\pi$-bond gets
more ionic under lattice expansion (as expected) whereas the Ti-O
$\sigma$-bond gets more covalent, which we will try to explain with
this model.

The parameters of this model can be extracted from the band energies
at symmetry points of the Brillouin zone in Fig.~\ref{Figure4} (see
the App.~C for more details).

For example, the on-site energies $\Delta_{m}$ can be obtained from
the $\Gamma$ point, since due to symmetry, the $d-p$ mixing vanishes
at this point and the band states acquire a pure $d$ or $p$ character.
For $a=3.8996$~{\AA} we have for STO $E_{d_{t2g}}\approx1.7$~eV,
$E_{d_{eg}}\approx4.3$~eV, $E_{p}\approx-1.2$~eV. This yields (using
Eq.~(\ref{ADelta_sigma}) and Eq.~(\ref{ADelta_pi}))
$2\Delta_{\pi}=E_{d_{t2g}}-E_{p}\approx2.9$~eV and
$2\Delta_{\sigma}=E_{d_{eg}}-E_{p}\approx5.5$~eV.

From these values and the $f_{mk}$ as given in
Refs.~\onlinecite{Wolfram73, Wolfram82} we obtain the Slater-Koster
hopping parameters $V_{\sigma}\approx2.1$~eV Eq.~(\ref{Vsigma}),
$V_{\pi}=V_{pd\pi}\approx1.6$~eV Eq.~(\ref{Vpi}) and
$V_{pd\sigma}\approx2.7$~eV Eq.~(\ref{Aeq:X1Hpt})\footnote{The
  parameters $V_{\pi}$ and $V_{\sigma}$ are from the WE ($p$-$d$)
  model and $V_{pd\pi}=V_{\pi}$ and $V_{pd\sigma}$ are from the
  Harrison ($s$-$p$-$d$) model, see App.~\ref{ap_STOBTO}.}.

Since the occupation numbers of the non-bonding bands do not depend on
the lattice and the anti-bonding bands ($\nu=+1$) are not occupied, we
consider the bonding bands ($\nu=-1$) only. The contributions from the
bonding bands to the population of the $p_m$ orbitals $n_{p_{m}}$ are
obtained by a sum over the Brillouin zone. In order to analyze the
occupation in dependence of the lattice expansion, we need the
derivative of the occupation number with respect to the lattice
parameter $a$.  From Eq.~(\ref{eq:Nknu}) we obtain for the derivative
(denoted by $\prime$)
\begin{eqnarray}
n_{p\mathbf{k}m,-1}^{\prime}
=\frac{V_{m}^{2}\left(f_{m\mathbf{k}}\right)^{2}\Delta_{m}}
{\left(\sqrt{\Delta_{m}^{2}+V_{m}^{2}\left(f_{m\mathbf{k}}\right)^{2}}\right)^{3}}
\left(\frac{\Delta_{m}^{\prime}}{\Delta_{m}}
-\frac{V_{m}^{\prime}}{V_{m}}\right).
\label{eq:Nprime}
\end{eqnarray}
The derivative of $n_{p_{m}}$ is proportional to
\begin{eqnarray}
n_{p_{m}}^{\prime}\propto\left(\frac{\Delta_{m}^{\prime}}
{\Delta_{m}}-\frac{V_{m}^{\prime}}{V_{m}}\right).
\label{eq:Npprime}
\end{eqnarray}
Fig.~\ref{Figure3} shows that $n_{p_{m}}^{\prime}$ has a different
behavior for $m=\sigma$ ($n_{p_{\sigma}}^{\prime}$ is negative) and
$m=\pi$ ($n_{p_{\pi}}^{\prime}$ is positive). Thus, within the WE model, the 
 observed increase of the EFG, which is due to the decreasing occupation of 
the ${p_{\sigma}}$-orbitals would yield
\begin{eqnarray}
-\frac{V_{\sigma}^{\prime}}{V_{\sigma}}<
-\frac{\Delta_{\sigma}^{\prime}}{\Delta_{\sigma}}.
\label{eq:cond}
\end{eqnarray}

Both $\Delta_{\sigma}$ and $V_{\sigma}$ decrease upon lattice
expansion: Fig.~\ref{Figure4} shows that the energies at the
$\Gamma$-point $E_{d_{eg}}$ and $E_{d_{t2g}}$ and the bandwidths are
smaller for the larger lattice parameter $a=4.009$~\AA, than for the
smaller lattice parameter $a=3.8996$~\AA.  A commonly accepted
estimate \cite{Harrison} for the dependence of hopping integrals on
$a$ is $V_{\sigma}\propto a^{-\alpha}$ with $\alpha$ between 3.5 and 4
(from the LDA band structure, we obtain $\alpha=3.5\pm0.5$).
This gives 
\begin{eqnarray}
-a\frac{V_{\sigma}^{\prime}}{V_{\sigma}}=\alpha\geq3.
\label{LHS}
\end{eqnarray}
$\Delta_{\sigma}$ is the difference in energy of the atomic levels
corrected by the crystal field (CF)\footnote{$\varepsilon$ denotes the
  energy of the atomic level and $E$, as used before, denotes the
  energy level corrected by the crystal field:
  $\Delta_{\sigma}=E_d-E_p =
  \varepsilon_{d}-\varepsilon_{p}+\delta_{CF,\sigma}$,
  cf. Eq.~(\ref{ADelta_sigma}). Note, that $\delta_{CF,m}$ is
  different for $m=\pi$ and $m=\sigma$, since $\varepsilon_{d}$ is the
  atomic energy level, and thus does not depend on $m$. This is the
  main reason, that, $\delta_{CF,\sigma}>\delta_{CF,\pi}$.
  Furthermore, $\delta_{CF,\sigma}$ has strong dependence on $a$.}
$\Delta_{\sigma}=\varepsilon_{d}-\varepsilon_{p}+\delta_{CF,\sigma}$.
The crystal field consists of two contributions \cite{crystalField}: A
(dominating) electrostatic contribution, which is the difference of
the Madelung potentials of Ti and O, hence $\delta_{CF,el}\propto
a^{-1}$, and a hybridization contribution, which, in our case
(octahedral coordination), contains a large and strongly $a$-dependent
contribution for $m=\sigma$ from the semi-core $s$-states of the
ligand. Indeed, (cf.~Fig.~\ref{Figure4}) the change due to the
increasing lattice parameter $a$ is much larger for $\Delta_{\sigma}$
than for $\Delta_{\pi}$. The main electrostatic contribution, which
implies $\delta_{CF,el}\propto a^{-1}$, leads to
\begin{eqnarray*}
-a\frac{\Delta_{\sigma}^{\prime}}{\Delta_{\sigma}}
=\frac{\delta_{CF,el}}{\Delta_{\sigma}}.
\end{eqnarray*}
Since $\varepsilon_{d}-\varepsilon_{p}+\delta_{CF,el}>\delta_{CF,el}$ is
$\delta_{CF,el}/\Delta_{\sigma}<1$ and therefore 
\begin{eqnarray}
  -a\frac{\Delta_{\sigma}^{\prime}}{\Delta_{\sigma}}<1.
\label{RHS}
\end{eqnarray}
Combining the estimates from Eq.~(\ref{LHS}) and Eq.~(\ref{RHS}), we
get
\begin{eqnarray}
  -a\frac{\Delta_{\sigma}^{\prime}}{\Delta_{\sigma}}<1<3
\leq-a\frac{V_{\sigma}^{\prime}}{V_{\sigma}}.
\label{wrongagain}
\end{eqnarray}
This is in contradiction to the inequality~(\ref{eq:cond}), leading to
the conclusion that the WE model, though consistent with the intuitive
expectations (see Sec.~\ref{Far}) is unable to predict the observed
behavior of the $\sigma$-orbital occupation in Fig.~\ref{Figure3}.

A possible reason for the failure of the WE model is that according to
Ref.~\onlinecite{Matth70}, a large contribution to the CF comes from
the oxygen 2$s$-orbitals, which lie almost 18~eV below the Ti 3$d$
level, $\Delta_{sd}=17.9$~eV, but have a large matrix element
$V_{sd\sigma}=3.0$~eV with the $e_{g}$ orbitals.  This suggests to
extend the WE model taking into account the oxygen 2$s$-states in
order to explain the increasing EFG upon lattice expansion. This is
Harrison's model, where $V_{sd\sigma}$ is obtained from
Eq.~(\ref{Aeq:G12H})
\begin{eqnarray*}
\Gamma_{12}=\frac{\varepsilon_{s}+\varepsilon_{d}}{2}
\pm\sqrt{(\frac{\varepsilon_{s}-\varepsilon_{d}}{2})^{2}+6V_{sd\sigma}^{2}},
\end{eqnarray*}
with $\varepsilon_{s}=-16.2$~eV, $\varepsilon_{d}=1.7$~eV
and $\Gamma_{12}=4.3$ taken from the band structure.

Taking the $s$-orbitals into account, $V_{\sigma}$ in the
inequality~(\ref{eq:cond}) is replaced by $V_{sd\sigma}$.
%left hand side of ineq.(\ref{eq:cond}) does not change, 
Harrison \cite{Harrison} argues that the $a$ dependence of
$V_{sd\sigma}$ is similar to the a dependence of $V_{pd\sigma}$.  This
suggestion is confirmed by our LDA calculations. Thus, we obtain
 \begin{eqnarray}
\frac{V_{sd\sigma}\prime}{V_{sd\sigma}}=-\frac{\alpha}{a}.
\label{LHS2}
\end{eqnarray}
On the right hand side we have the on-site energy difference, which is
given by
$\Delta_{\sigma}\approx\Delta_{\pi}+3V_{sd\sigma}^{2}/\Delta_{sd}$,
cf. Eq.~(\ref{ADeltaSigma}). The derivative of this expression is
\begin{eqnarray}
\Delta_{\sigma}^{\prime}\approx\frac{6}{\Delta_{sd}}V_{sd\sigma}V_{sd\sigma}^{\prime}.
\label{RHS2}
\end{eqnarray}
Note, that here we assumed $\Delta_{\pi}^{\prime}
=\Delta_{sd}^{\prime}=0$.  Applying Eqs.~(\ref{RHS2}) and (\ref{LHS2})
yields
\begin{eqnarray}
-\frac{\Delta_{\sigma}^{\prime}}{\Delta_{\sigma}}
=\frac{\alpha}{a}\frac{6V_{sd\sigma}^{2}}{\Delta_{sd}\Delta_{\pi}+
  3V_{sd\sigma}^{2}}.
\label{RHS3}
\end{eqnarray}
Inserting Eq.~(\ref{LHS2}) and Eq.~(\ref{RHS3}) in the
inequality~(\ref{eq:cond}), we obtain within the Harrison model the
observed increase of the EFG, due to the decreasing occupation of the
${p_{\sigma}}$-orbitals, if the following inequality is fulfilled:
\begin{eqnarray}
\frac{\alpha}{a}=-\frac{V_{sd\sigma}\prime}{V_{sd\sigma}} & \stackrel{!}{<} & 
-\frac{\Delta_{\sigma}^{\prime}}{\Delta_{\sigma}}
=\frac{\alpha}{a}\frac{6V_{sd\sigma}^{2}}
{\Delta_{sd}\Delta_{\pi}+3V_{sd\sigma}^{2}}
\nonumber 
\\
\Leftrightarrow\quad  \frac{1}{3}\Delta_{sd}\Delta_{\pi} & \stackrel{!}{<} & 
V_{sd\sigma}^{2}
.
\label{correct}
\end{eqnarray}

Using the values obtained from the LDA band structure
($V_{sd\sigma}=3.0$~eV, $\Delta_{sd}=17.9$~eV and $\Delta_{\pi}=1.4$),
we see that Eq.~(\ref{correct}) is fulfilled.
Thus, the inequality Eq.~(\ref{eq:cond}) holds for the STO $\sigma$-orbitals
and the observed negative slope of $n_{z}$ in Fig.~\ref{Figure3}
can be understood.

After revealing the origin of the counter-intuitive behavior of the
on-site EFG, we will discuss the unusually large value of the off-site
EFG of the considered compounds.
The dependence of this contribution with respect to the lattice
parameter can be estimated in the following way: From the multipole
expansion of a potential of a given ion, the sum of the monopole
contributions to $v^{off}(\mathbf{r})$ Eq.~(\ref{eq:voff}) has the
slowest convergence. This contribution may be calculated within a
point charge model (PCM). Therefore, we note that the $V_{zz}$ value
created in the origin by a unit charge situated at the point
$\mathbf{R}$ equals the value of the $z$-component of the electric
field $E_z$, created in the origin by the unit dipole directed along
$z$-axis and situated at the same point $\mathbf{R}$:
$V_{zz}=(3Z^2-R^2)/R^5$.
That means, for the calculation of the EFG within the PCM, we need the
electric field $S(\mathbf{r})$ of dipoles located at the sites
$\mathbf{R}$, which are polarized along the $z$ direction and whose
polarization is unity, at various points $\mathbf{r}$ through the
cubic lattice: $S(\mathbf{r})=\sum_{\mathbf R} E_z(\mathbf R - \mathbf
r)$.  Here, $\mathbf r = a(x,y,z)$ and $\mathbf R= a(l,m,n)$ with $a$ being
the lattice parameter and $l,m,n =0,\pm 1,\pm 2$.
Using Eq.~(16) of Ref.~\onlinecite{Slater50}, we obtain for the EFG in
the PCM at the oxygen site
\begin{eqnarray}
V^{PCM}_{zz} &=& -\frac{e}{a^3}\left[n_{Ti}S(0,0,\frac12)
+n_{A}S(\frac12,\frac12,0)\right. \nonumber \\
& &\left. +2n_{O}S(0,\frac12,\frac12)\right] \nonumber \\
&=& -\frac{e}{a^3}\left[30.080n_{Ti}-
8.668\left(n_{A}-n_{O}\right)\right]. \label{PCM}
\end{eqnarray}
Here, $n_{Ti}$ is the monopole moment of the ionicity of Ti. 
If we insert the charges of the Ti ion $n_{\rm{Ti}}$, the O ion
$n_{\rm O}$, and the A ion $n_{\rm A}=-(n_{\rm{Ti}}+3n_{\rm O})$ (with
A=Sr, Ba) obtained from the FPLO calculations, we obtain e.g. for STO
$V^{PCM}_{zz}=1.30\cdot10^{21}$~V/m$^2$. This value is very close to
$V_{zz}^{off}=1.19\cdot10^{21}$~V/m$^2$, see Tab.~\ref{Tab1}. So, we
obtain a good agreement for the EFGs obtained from the simple PCM
model and the more complex calculation. This means, the FPLO code
yields realistic relations of the charge distributions.

The prefactor ${e}/{a^3}$ in Eq.~(\ref{PCM}) is responsible for
the observed decrease of the off-site contribution in case of lattice
expansion, see Fig.~\ref{Figure2}. Also the charge redistribution may
change the value of $V_{zz}^{off}$, but as we see in
Fig.~\ref{Figure2}, it has a minor effect: The off-site EFG for BTO is
smaller than for STO, but the distance between the two curves is
smaller than the lattice parameter dependence of the two curves.

\section{Summary and Conclusion}

In summary, we have performed first principle calculations of the
electric field gradient on the oxygen site for BaTiO$_{3}$ and
SrTiO$_{3}$ for different lattice parameters $a$. The values of our
calculated EFGs agree well with the measured and, apart from the sign,
with the calculated (LAPW) counterparts from
Ref.~\onlinecite{Blinc08}.

Decomposition of the EFG yields a large on-site contribution
originating from the oxygen 2$p$ shell. The on-site EFG reveals an
anomalous dependence of the $p_\sigma$-orbital population with respect
to the lattice parameter $a$: The population decreases under lattice
expansion, i.e. the $p$-$d$ hybridization grows with increasing Ti-O
distance. Simple ionic and covalent approaches lead to the conclusion
that this behavior is counter-intuitive. Also the effective two-level
Hamiltonian proposed by Wolfram and Ellialtioglu, which describes the
relevant states of the valence region (oxygen $p$- and titanium
$d$-states) fails to describe the observed behavior of the EFG upon
lattice expansion.  Only the inclusion of the O 2$s$ states to the
crystal field results in a consistent picture:
In fact, lattice expansion causes a charge transfer from the
$p_\sigma$- to the $s$-orbitals of oxygen, whereas the population of
the oxygen $\pi$-orbitals increases with $a$. This charge
redistribution leads to the increase of the EFG, which is the main
reason for the surprisingly large difference of the EFGs between
BaTiO$_{3}$ and SrTiO$_{3}$.

We expect that the observed feature, the increase of the anisotropy
count of the $p$-shell with the bond length, is common to all
$d$-metal-oxygen bonds and should be taken into account accordingly in
the interpretation of the relevant experiments.

The considered $A$TiO$_{3}$ systems are not strongly correlated, since
the Ti 3$d$ shell is formally empty. For magnetic ions with partially
filled $d$-shells, the influence of the O 2$s$ orbitals will be
diminished because the charge transfer energy $\Delta_{sd}$ will
include the on-site Coulomb repulsion within the $d$-shell.

As a side effect, our investigation sounds a a note of caution: When
performing a mapping of a complex DFT band structure calculation onto
a microscopically based minimal model in order to gain deeper physical
understanding, care has to be taken that all relevant interactions
are included.

\section*{Acknowledgments}

The authors thank the Heisenberg-Landau, "DNIPRO" (14182XB) and the
SPP 1178 of the Deutsche Forschungsgemeinschaft programs for
support. Discussions with V.\ V.\ Laguta and Alim Ormeci are
gratefully acknowledged.

\section{Appendix}

\subsection{EFG implementation in FPLO}\label{appA}

The EFG is a local property. It is a traceless symmetric tensor of
rank two, defined as the second partial derivative of the potential
$v(\mathbf r)$ evaluated at the position of the nucleus
\begin{eqnarray}
\label{EFGcart}
V_{ij} &\equiv&\left( \frac{\partial^2 v(\mathbf r)}{\partial_i \,
  \partial_j} -\frac{1}{3}\delta_{ij}\Delta v(\mathbf
r)\right)\Bigg|_{\mathbf r=0}.
\end{eqnarray}
With the definition
\begin{eqnarray}
\label{V2mdef}
V_{2m}&\equiv&\sqrt{\frac{15}{4 \pi}} \lim_{r\rightarrow
  0}\frac{1}{r^2}v_{2m}(r),
\end{eqnarray}
can the Cartesian EFG tensor Eq.~(\ref{EFGcart}) also be expressed in
(real) spherical components ($l=2$, $m=\pm2,\pm1,0$)
\begin{eqnarray}
\label{EFGtensor}
V_{ij}=
%\nonumber
%&&
\left(
\begin{array}{ccc}
 V_{22} -\frac{1}{\sqrt{3}}V_{20}  &  V_{2,-2}  
&  V_{21}  
\\
  V_{2,-2}  &   -V_{22} -\frac{1}{\sqrt{3}}V_{20}  
& V_{2,-1} 
\\
  V_{21}  &   V_{2,-1}     &     \frac{2}{\sqrt{3}}V_{20} 
\end{array}
\right).
\end{eqnarray}
In FPLO, the EFG on a nucleus at a given lattice site $\mathbf{s}_{0}$
may be represented as the sum of two contributions
\begin{eqnarray}
\label{eq:Vij}
V_{ij} & \equiv & 
\left(\frac{\partial^{2}}{\partial_i\partial_j}
-\frac{1}{3}\delta_{ij}\Delta \right)
\left[v^{on}(\mathbf{r}\mathbf{)}+v^{off}(\mathbf{r)}\right]
%=V_{ij}^{on}+V_{ij}^{off}
\\
v^{on}(\mathbf{r}\mathbf{)} & = & 
\sum_{L}\int d^{3}\mathbf{r}^{\prime}
\frac{n_{\mathbf{s_0},L}\left(\left|\mathbf{r}^{\prime}\right|\right)Y_{L}
\left(\mathbf{r}^{\prime}\right)}{\left|\mathbf{r}-\mathbf{s}_{0}-
\mathbf{r}^{\prime}\right|},\label{eq:von}
\\
v^{off}(\mathbf{r)} & = & 
\sum_{\mathbf{R}+\mathbf{s}\neq\mathbf{s}_{0},L}
\int d^{3}\mathbf{r}^{\prime}\frac{n_{\mathbf{s},L}
\left(\left|\mathbf{r}^{\prime}\right|\right)Y_{L}
\left(\mathbf{r}^{\prime}\right)}
{\left|\mathbf{r}-\mathbf{R}-\mathbf{s}
-\mathbf{r}^{\prime}\right|}\label{eq:voff}
\\
 & - & \sum_{\mathbf{R}+\mathbf{s}\neq\mathbf{s}_{0}}
\frac{Z_{\mathbf{s}}}{\left|\mathbf{r}-\mathbf{R}-\mathbf{s}\right|},
\nonumber
\end{eqnarray}
where $Y_{L}$ are the (real) spherical harmonics; $\mathbf{R}$ is a
Bravais vector, and $\mathbf{s}$ is an atom position in the unit
cell. The index $L=nlm$ also absorbs the spin and the principal
quantum number.  The first term in Eq.~(\ref{eq:Vij}), the on-site
contribution, comes from the on-site contribution of the electron
density of the site $\mathbf s_0$, and the second term, the off-site
contribution, comes from the potential of all other atoms.

Since the angular momentum components of the local charge density give
rise to multipole moments, which determine the Coulomb potential for
large distances, FPLO uses the Ewald method to handle the long-range
interactions (see \cite{FPLO} section D). The density is modified
with a Gaussian auxiliary density $\tilde
n_l(r)=n_l(r)-n_l^{Ew}(r)$\footnote{Note, that we use another sign in
  the definition of $n^{Ew}$ compared to \cite{FPLO} Eq (46) and
  (47).}. Inserting this modified density in the potentials
Eq.~(\ref{eq:von}) and Eq.~(\ref{eq:voff}) yields
\begin{eqnarray}
\label{sum}
v(\mathbf r)%=v^{on}(\mathbf r)+v^{off}(\mathbf r)
=\tilde v^{on}(\mathbf r)+v^{Ew,on}(\mathbf r)
+\tilde v^{off}(\mathbf r) +v^{Ew,off}(\mathbf r).
\end{eqnarray}
These contributions are calculated to get the total EFG.

The first contribution is $\tilde v^{on}(\mathbf r)$ in
Eq.~(\ref{sum}). This potential is given by Eq.~(\ref{eq:von}) using the
modified density $\tilde n_{\mathbf{s_0},L}(r')$. The corresponding
$\tilde v_{\mathbf{s_0},2m}(r)$ components needed in
Eq.~(\ref{V2mdef}) are obtained from the solution of the radial
Poisson equation (see Ref.~\onlinecite{FPLO} Eq.~(49))
\begin{eqnarray*}
\tilde v_{\mathbf{s_0},L}(r)&=&
\frac{4\pi}{2l+1}\Big[\frac{1}{r^{l+1}}\int_0^r dx x^{l+2}\tilde
  n_{\mathbf{s_0},L}(x) 
\\ & & +r^l\int_r^\infty dx x^{-l+1}\tilde
  n_{\mathbf{s_0},L}(x)\Big].
\end{eqnarray*}
Using the rule of L'Hospital we obtain for the $\tilde V^{on}_{2m}$
component (from which $\tilde V^{on}_{ij}$ is obtained) 
\begin{eqnarray}
\label{v2m/r^2}
\tilde V^{on}_{2m}=
2\sqrt{\frac{3\pi}{5}}\left[\frac{n_{\mathbf{s_0},2m}(0)}{5} 
+\int_0^\infty \!\!\!\! dx x^{-1}\tilde n_{\mathbf{s_0},2m}(x)\right]\
\end{eqnarray}
The first term in Eq.~(\ref{v2m/r^2}) is the $2m$ component of the
electronic density at the nucleus $n_{\mathbf{s_0},2m}(0)\equiv \tilde
n_{\mathbf{s_0},2m}(0)$.  The $n_{2m}$ component of a spherical
harmonic expansion of an analytic function around a given point
behaves as $n_{2m}=\mathcal{O}(r^2)$. The only non-analyticities of
the electron density are caused by the spherical singularities of the
nuclear potential and this can not be aspherical. Therefore
$n_{2m}(0)=0$, which can be shown explicitly both in a
non-relativistic and full relativistic theory.

The second contribution is $\tilde v^{off}(\mathbf r)$ in
Eq.~(\ref{sum}). This potential is given by Eq.~(\ref{eq:voff}) using
the modified density $\tilde n_{\mathbf{s},L}(r')$.  Since the density
$n_{\mathbf s,2m}$ is not given at the site $\mathbf s_0$, where the
atom under consideration is sitting, this equation has to be
expanded. This can be done explicitly but the derivation as well as
the result for $\tilde V_{ij}^{off}$ are very bulky~\cite{KKphdthesis}
and therefore not given here.

The third contribution are $v^{Ew,on}(\mathbf r)+v^{Ew,off}(\mathbf
r)$ in Eq.~(\ref{sum}), which have to be calculated from the Ewald
density alone. The auxiliary density $n_l^{Ew}(\mathbf r)$ is given as
a Fourier expansion, resulting in the Ewald potential in Fourier space
$v_{\mathbf G}^{Ew}=\frac{4\pi}{|\mathbf G|^2}n_{\mathbf G}^{Ew}$,
Eq.~(52) in \cite{FPLO}. $V_{ij}^{Ew}$ is obtained by differentiating
$v^{Ew}(\mathbf r)=\sum_{\mathbf G}e^{i\mathbf G \mathbf s}v_{\mathbf
  G}^{Ew}$
\begin{eqnarray}
V_{ij}^{Ew}=-\sum_{\mathbf G}
\left(G_iG_j-\frac{1}{3}\mathbf G^2 \delta_{ij}\right)
\Re(e^{i\mathbf G \mathbf s}v_{\mathbf G}^{Ew})
\label{V^Ew}
\end{eqnarray}

The total EFG tensor $V_{ij}$ is given by the sum of these three
contributions
\begin{eqnarray}
\label{totEFG}
V_{ij}=\tilde V_{ij}^{on}+\tilde V_{ij}^{off}+V_{ij}^{Ew}.
\end{eqnarray}
In order to analyze the on-site and off-site contributions, we define
the on-site EFG as being the first term in Eq.~(\ref{eq:Vij}), but
calculated from the unmodified density 
\begin{eqnarray}
\label{Vonsite}
V_{2m}^{on}=2\sqrt{\frac{3\pi}{5}}
 \int_0^\infty dx x^{-1} n_{\mathbf{s_0},2m}(x). \quad
\end{eqnarray}
The off-site EFG then is taken to be 
\begin{eqnarray}
\label{Voffsite}
V_{2m}^{off}= V_{2m}- V_{2m}^{on}.
\end{eqnarray}

\subsection{Orbital contributions to the EFG}
\label{appB}

In FPLO the electron density is separated into a net density and an
overlap density (see Ref.~\onlinecite{FPLO} section B). The dominating
net density is calculated from two orbitals at the same site $\mathbf
R+\mathbf s=\mathbf R'+\mathbf s'=\mathbf s_0$
\begin{eqnarray*}
n^{net}_{\mathbf s_0}(\mathbf r)=\!\!\!\! 
\sum_{\mathbf k,n\newline L_1,L_2}^{occ}
\!\!\!\! 
c_{\mathbf{s_0}L_1}^{\mathbf{k},n}
\varphi_{\mathbf{s_0},L_1}(\mathbf{r}-\mathbf{s_0})
\cdot
{c}_{\mathbf{s_0}L_2}^{\star\mathbf{k},n}
\varphi_{\mathbf{s_0},L_2}(\mathbf{r}-\mathbf{s_0}).
\end{eqnarray*}
The basis functions $\varphi_{\mathbf{s_0},L}$ are localized on the
lattice sites
 \begin{eqnarray*}
\varphi_{\mathbf{s_0},L}(\mathbf{r}-\mathbf{s_0})\equiv
\phi_{\mathbf{s_0}}^l(\left|\mathbf{r}-\mathbf{s_0}\right|)
Y_{L}\left(\mathbf{r}-\mathbf{s_0}\right).
\end{eqnarray*}
%and $c_{\mathbf{s_0}L}^{\mathbf{k},n}$ 
The $2m$ component of the radial net density, needed for the
contributions of the net EFG, can be calculated from
\begin{eqnarray}
\label{n2m}
n^{net}_{\mathbf{s_0},2m}(r)&=&
\int n^{net}_{\mathbf s_0}(\mathbf r) 
Y_{2m}\left(\mathbf{r}-\mathbf{s_0}\right)d\Omega
\\
&=&\!\!\!\! \sum_{ L_1,L_2}%^{occ}
\!\!\!
c_{L_1L_2}
\phi_{\mathbf{s_0}}^{l_1}(\left|\mathbf{r}-\mathbf{s_0}\right|)
\phi_{\mathbf{s_0}}^{l_2}(\left|\mathbf{r}-\mathbf{s_0}\right|)
G_{l_1,l_2,2}^{m_1,m_2,m},
\nonumber
\end{eqnarray}
where $G_{l_1,l_2,2}^{m_1,m_2,m}$ are the Gaunt coefficients and
$c_{L_1L_2}=\sum_{\mathbf{k},n}
c_{\mathbf{s_0}L_1}^{\mathbf{k},n}{c}_{\mathbf{s_0}L_2}^{\star\mathbf{k},n}$.
Due to the properties of the Gaunt coefficients,
$n^{net}_{\mathbf{s_0},2m}$ consists only of $p$-$p$, $d$-$d$, and
$s$-$d$ (and if present $p$-$f$ and $f$-$f$) contributions.  These
contributions to the on-site net EFG $V_{zz}^{on,net}$ are obtained by
inserting Eq.~(\ref{n2m}) into Eq.~(\ref{Vonsite}). E.g. the $p$-$p$
contribution $V_{2m,pp}^{on,net}$ is calculated from
\begin{eqnarray}
\label{eq:V2mpp}
V_{2m,pp}^{on,net}&=&2\sqrt{\frac{3\pi}{5}}
\int_0^\infty dx x^{-1} n^{net,pp}_{\mathbf{s_0},2m}(x)
\\
\nonumber
n^{net,pp}_{\mathbf{s_0},2m}(x)&=&
[\phi_{\mathbf{s_0}}^{1}(x)]^2
\!\!\!\! \sum_{m_1,m_2}%^{occ}
\!\!\!\!\!\!
c_{1,1}^{m_1,m_2}
G_{1,1,2}^{m_1,m_2,m}.
\end{eqnarray}
The main component $V_{zz,pp}^{on,net}=\frac{2}{\sqrt
  3}V_{20,pp}^{on,net}$ is calculated from
\begin{eqnarray}
\label{eq:Vzzpp}
n^{net,pp}_{\mathbf{s_0},20}(x)=
\sqrt{\frac{1}{5\pi}}
[\phi_{\mathbf{s_0}}^{1}(x)]^2
\!\sum_{\mathbf k,n}%^{occ}
\!
%\qquad \qquad \qquad \qquad \qquad \cdot
\Bigg( 
c_{\mathbf{s_0},1,0}^{\mathbf{k},n}{c}_{\mathbf{s_0},1,0}^{\star\mathbf{k},n}
\\
-\frac{1}{2}
\left(c_{\mathbf{s_0},1,-1}^{\mathbf{k},n}{c}_{\mathbf{s_0},1,-1}^{\star\mathbf{k},n}
+
c_{\mathbf{s_0},1,1}^{\mathbf{k},n}{c}_{\mathbf{s_0},1,1}^{\star\mathbf{k},n}\right)
\Bigg).
\nonumber
\end{eqnarray}
We see, that this density is proportional to the difference of
occupation in $p_z$ ($m=0$) and $p_{x,y}$ ($m=\pm1$) states, which is
the anisotropy count.

\subsection{Background for Sec.~\ref{disc}} \label{ap_STOBTO}

\begin{table}[t]
\begin{center}
\begin{tabular}{|l|r|r|r|r|r|r|r|r|r|}
\hline 
$a$  [\AA] & $\Gamma_{12}$  & $\Gamma_{1}$  & $\Gamma_{15}$  
& $\Gamma_{25}$  & $\Gamma_{15}$  
& $\Gamma_{25}^{\prime}$  
& $\Gamma_{12}$  & $X_{5}$  & $X_{1}$
\tabularnewline
\hline 
3.8996 & -17.199  & -16.177  & -2.891  & -1.166  & -0.372  & 1.709  & 4.319 &3.705  & 6.551 
\tabularnewline
\hline 
4.009  & -16.923  & -15.968  & -2.828  & -1.046  & -0.408 & 1.579  & 3.800  & 3.332  & 5.798 
\tabularnewline
\hline
\end{tabular}
\end{center}
\caption{The energies at the $\Gamma$ and $X$ points in SrTiO3 given in
  eV. Here, $\Gamma_{1}\approx\varepsilon_{s}$,
  $\Gamma_{25}\approx\varepsilon_{p}$,
  $\Gamma_{25}^{\prime}=E_{d_{t2g}}\approx\varepsilon_{d}$ and
  $\Gamma_{12}=E_{d_{eg}}$}
\label{Atab:TabE}
\end{table}

In order to extract the parameters from the band structure we need the total
Hamiltonian 
\begin{eqnarray}
\hat{H}=\sum_{m}
\left[\hat{H}_{m}+e_{m}
\left(\left|d,\mathbf{k}\right\rangle \left\langle d,\mathbf{k}\right|
+\left|p,\mathbf{k}\right\rangle \left\langle
p,\mathbf{k}\right|\right)\right].
\label{Aeq:HWE}
\end{eqnarray}
Here, $\hat{H}_{m}$ is the Hamiltonian given in
Eq.~(\ref{eq:Htlsproj}) and $e_{m}$ is the mean energy of a pair of
bands. The energies are therefore obtained from
\begin{eqnarray}
\label{energyEV}
E_{\mathbf{k}m\nu} + e_{m} 
= e_m +  \nu\sqrt{\Delta_{m}^{2}+V_{m}^{2}\left(f_{m\mathbf{k}}\right)^{2}}.
\end{eqnarray}
For the three pairs of the $\pi_{ij}$ bands, $f_{m\mathbf{k}}$ is given by
\begin{eqnarray}
f_{\pi_{ij}\mathbf{k}}^{2}=2\left(2-C_{i}-C_{j}\right)\quad ,
\label{Aeq:piab}
\end{eqnarray}
with $C_{i}\equiv\cos(k_{i}a)$.  The two $\sigma$-bands are distinguished by
the index $\lambda=\pm1$ and $f_{m\mathbf{k}}$ is
\begin{eqnarray}
\nonumber
f_{\sigma_{\lambda}\mathbf{k}}^{2}
=3-C_{x}-C_{y}-C_{z} \qquad \qquad\qquad \qquad\qquad \qquad
\\ 
+\lambda\left(C_{x}^{2}+C_{y}^{2}+C_{z}^{2}-C_{x}C_{y}-C_{x}C_{z}-C_{y}C_{z}
\right)^{1/2}.
\label{Aeq:sgml}
\end{eqnarray}
Inserting these in Eq.~(\ref{energyEV}) for the $\Gamma$ point
($\mathbf{k}a=0$), and the $X$ point, ($k_{x}a=\pi$, $k_{y}=k_{z}=0$) we obtain
\begin{eqnarray} 
\Gamma_{25} &=& e_{\pi}-\Delta_{\pi}=e_{\sigma}-\Delta_{\sigma},
\label{Aeq:G25} 
\\
\nonumber
\mbox{and} \quad \Gamma_{25} &\equiv& E_{p}  \approx \varepsilon_{p}
\\
\Gamma_{25}^{\prime} &=&e_{\pi}+\Delta_{\pi},
\label{Aeq:G25p}
\\
\nonumber
\mbox{and} \quad
\Gamma_{25}^{\prime} &\equiv& E_{d_{t2g}} \approx \varepsilon_{d}
\\ 
\Gamma_{12} & =& 
e_{\sigma}+\Delta_{\sigma},
\label{Aeq:G12}
\\
\nonumber
 \mbox{and} \quad
\Gamma_{12} &\equiv& E_{d_{eg}}
\\ 
X_{5} & = &
e_{\pi}+\sqrt{\Delta_{\pi}^{2}+4V_{\pi}^{2}},
\label{Aeq:X5}
\\ 
X_{1} & = &
e_{\sigma}+\sqrt{\Delta_{\sigma}^{2}+4V_{\sigma}^{2}}.
\label{Aeq:X1}
\end{eqnarray}
Here, $\varepsilon$ denotes the energy of the atomic level, and $E$ denotes
the energy level corrected by a 'crystal field' $\delta_{CF}$, see below.

Now it is trivial to find the parameters $\Delta_{m},V_{m}$ 
\begin{eqnarray}
2\Delta_{\pi} & = & \Gamma_{25}^{\prime}-\Gamma_{25}
\equiv E_{d_{t2g}}-E_{p},
\label{ADelta_pi}
\\ 
2\Delta_{\sigma} & = & \Gamma_{12}-\Gamma_{25}
\equiv E_{d_{eg}}-E_{p}=\varepsilon_d -\varepsilon_p +\delta_{CF} ,
\label{ADelta_sigma}
\\ 
4V_{\pi}^{2} & = &
\left(X_{5}-\Gamma_{25}^{\prime}\right)\left(X_{5}-\Gamma_{25}\right),
\label{Vpi}
\\ 
4V_{\sigma}^{2}  & = &
\left(X_{1}-\Gamma_{12}\right)\left(X_{1}-\Gamma_{25}\right).
\label{Vsigma}
\end{eqnarray}
The energy values at the different $\Gamma$ and $X$ points for SrTiO$_3$ are
given Tab.~\ref{Atab:TabE}.

So far, we have used the WE model, i.e.  we have taken into account only the
oxygen $p$ and the titanium $d$ states. Since this model is not sufficient to
explain the observed behaviour of the oxygen $p_\sigma$-states, we have to
expand the model.  Harrison's model \cite{Harrison} includes also the oxygen
$s$-states.
%(and $p-p$ hopping also, which I have neglected), 
The $s$-states change the dispersion in the $\sigma$ bands, so that we have
two parameters $V_{pd\sigma}$, $V_{sd\sigma}$ instead of just one
$V_{\sigma}$. Thus, the expressions become more complex, even in the symmetry
points. In this model, the Eqs.~(\ref{Aeq:G25}), (\ref{Aeq:G25p}) and
(\ref{Aeq:X5}) remain the same and the parameters $\varepsilon_{p}$,
$\varepsilon_{d}$ and $V_{\pi}$ are unchanged. For $\Gamma_{12}$
Eq.~(\ref{Aeq:G12}) and $X_1$ Eq.~(\ref{Aeq:X1}), Harrison obtains
\begin{eqnarray} 
\Gamma_{12} & = &
\frac{\varepsilon_{d}+\varepsilon_{s}}{2}
+\sqrt{\left(\frac{\varepsilon_{d}-\varepsilon_{s}}{2}\right)^{2}+6V_{sd\sigma}^{2}},
\label{Aeq:G12H}
\\ 
X_{1}&\approx&\frac{\varepsilon_{d\sigma}+\varepsilon_{p}}{2}
+\sqrt{\left(\frac{\varepsilon_{d\sigma}-\varepsilon_{p}}{2}\right)^{2}
+4V_{pd\sigma}^{2}},
\label{Aeq:X1Hpt}
\end{eqnarray}
where $\varepsilon_{d\sigma}=\varepsilon_{d}+2V_{sd\sigma}^{2}/\Delta_{sd}$.
From these equations, the parameters $V_{pd\sigma}$ and $V_{sd\sigma}$ can be
obtained. Besides, there is also an additional equation for $\Gamma_{1}$
\begin{eqnarray}
\Gamma_{1}=\varepsilon_{s}.
\label{Aeq:G1}
\end{eqnarray}

Substituting $\Delta_{sd}\equiv\varepsilon_{d}-\varepsilon_{s}\gg
V_{sd\sigma}$ in Eq.~(\ref{Aeq:G12H}), we obtain
\begin{eqnarray*}
\Gamma_{12} & = & \frac{\varepsilon_{d}+\varepsilon_{s}}{2}
+\left(\frac{\Delta_{sd}}{2}\right)\sqrt{1+\frac{24V_{sd\sigma}^{2}}{\Delta_{sd}^{2}}}
\\
 & \approx & \frac{\varepsilon_{d}+\varepsilon_{s}}{2}
+\left(\frac{\Delta_{sd}}{2}\right)
\left[1+\frac{12V_{sd\sigma}^{2}}{\Delta_{sd}^{2}}\right]
\\
 & = & \varepsilon_{d}+6\frac{V_{sd\sigma}^{2}}{\Delta_{sd}}
\end{eqnarray*}
 Hence, 
\begin{eqnarray}
E_{d_{eg}}\equiv\Gamma_{12}\approx\varepsilon_{d}+\frac{6V_{sd\sigma}^{2}}{\Delta_{sd}},
\label{Aeq:G12Hpt}
\end{eqnarray}

For the main text, we need an expression for $\Delta_{\sigma}$:
\begin{eqnarray}
\nonumber
\Delta_{\sigma}&=&\left(\Gamma_{12}-\Gamma_{25}\right)/2
\\
\nonumber
&=&\left(\Gamma_{12}-\varepsilon_{p}\right)/2
\\
\nonumber
&\approx&\frac{1}{2}\left(\varepsilon_{d}+\frac{6V_{sd\sigma}^{2}}{\Delta_{sd}}
-\varepsilon_{p}\right)
\\
&=&\Delta_{\pi}+3V_{sd\sigma}^{2}/\Delta_{sd}.
\label{ADeltaSigma}
\end{eqnarray}

Finally the hopping parameters of both models are given in the Table
\ref{Atab:TabV}

\begin{table}
\begin{center}
\begin{tabular}{|c|c|c|c|c|}
\hline 
$a$ & $V_{sd\sigma}$ & $V_{pd\sigma}$ & $V_{\sigma}$ & $V_{pd\pi}=V_{\pi}$\tabularnewline
\hline 
3.9 & 2.9855 & 2.7237 & 2.0754 & 1.5590\tabularnewline
\hline 
4.0 & 2.7054 & 2.4064 & 1.8486 & 1.3854\tabularnewline
\hline
\end{tabular}
\end{center}
\caption{parameters of WE and Harrison models}
\label{Atab:TabV}
\end{table}

\medskip

Remark:

\smallskip

In the WE model, we use $E_m$ as model parameter, hence $\Gamma
\approx \varepsilon$, and in the Harrison model, we use
$\varepsilon_m$ as model parameter, hence $\Gamma = \varepsilon$.
However. there is some contribution of the CF acting on the
e.g. $p$-states at the $\Gamma$ point: The interactions with Sr
states, with core states, with Madelung potentials etc. Therefore,
$\varepsilon_p$ is rather a model parameter than the true atomic
energy, $E_p$, of a $2p$ state.  If we speak about the model only, we
may drop $E_p$ and $E_{t2g}$, and retain only $\varepsilon_p$,
$\varepsilon_d$ and $E_{eg}$.

\end{document}